# Survivable Routing in IP-over-WDM Networks in the Presence of Multiple Failures

Maciej Kurant, Patrick Thiran
LCA - School of Communications and Computer Science
EPFL, CH-1015 Lausanne, Switzerland
Email: {maciej.kurant, patrick.thiran}@epfl.ch

*Abstract*— **Failure restoration at the IP layer in IP-over-WDM networks requires to map the IP topology on the WDM topology in such a way that a failure at the WDM layer leaves the IP topology connected. Such a mapping is called** *survivable*. **As finding a survivable mapping is known to be NP-complete, in practice it requires a heuristic approach. We have introduced in [1] a novel algorithm called "SMART", that is more effective and scalable than the heuristics known to date. Moreover, the formal analysis of SMART [2] has led to new applications: the formal verification of the existence of a survivable mapping, and a tool tracing and repairing the vulnerable areas of the network. In this paper we extend the theoretical analysis in [2] by considering** *multiple failures*.

## I. INTRODUCTION

Generally, there are two approaches for providing survivability of IP-over-WDM networks: protection and restoration [3]. Protection uses pre–computed backup paths applied in the case of a failure. Restoration finds dynamically a new path, once a failure has occurred. Protection is less resource efficient (the resources are committed without prior knowledge of the next failure) but fast, whereas restoration is more resource efficient and slower. Protection and restoration mechanisms can be provided at different layers. *IP layer* (or *logical* layer) survivability mechanisms can handle failures that occur at both layers, contrary to *WDM layer* (or *physical* layer) mechanisms that are transparent to the IP topology. It is not obvious which combination (mechanism/layer) is the best; each has pros and cons [4]. IP restoration, however, deployed in some real networks, was shown to be an effective and cost–efficient approach (see e.g., Sprint network [5]). In this paper we will consider exclusively the *IP restoration approach*.

Each logical (IP) link is mapped on the physical (WDM) topology as a *lightpath*. Usually a fiber is used by more than one lightpath (in Sprint the maximum number is 25 [6]). Therefore, even a single physical link failure usually brings down a number of IP links. With the IP restoration mechanism, these IP link failures are detected by IP routers, and alternative routes in the IP topology are found. In order to enable this, the IP topology should remain *connected* after failures; this in turn may be guaranteed by an appropriate mapping of IP links on the physical topology. Such a mapping is called a *survivable mapping*.

For a given pair of physical an logical topologies, finding a survivable mapping is an NP-complete problem [7]. Therefore the exact approaches, such as Integer Linear Programming [7], [8], do not scale well. For this reason various heuristics were proposed, e.g., Tabu Search [8], [9], [10], Simulated Annealing [4] and others [3], [11]. In [1] we have proposed a novel approach that led us to a heuristic algorithm called "SMART", that is much more effective and scalable than the heuristics known to date.

The SMART algorithm, however, is not only a heuristic. The theoretical studies in [2] have revealed a number of useful properties of our algorithm. This was made possible by the introduction of a new type of mapping that preserves the survivability of some subgraphs ('pieces') of the logical topology; we call it a *piecewise survivable mapping*. The formal analysis of the piecewise survivable mapping shows that a survivable mapping of the logical topology on the physical topology exists if and only if there exists a survivable mapping for a *contracted* logical topology, that is, a logical topology where a specified subset of edges is contracted (contraction of an edge amounts to removing it and merging its end-nodes). This result substantially simplifies the verification of the existence of a survivable mapping, making it, for the first time, often possible for moderate and large topologies. A second application of a piecewise survivable mapping is tracing the vulnerable areas in the network and pointing where new link(s) should be added to enable a survivable mapping [2].

This paper extends the theoretical results in [2] by considering *multiple failures*, i.e., independent failures of a number of physical links. Usually such a situation takes place when a failure occurs before another one is repaired. This is possible in practice. For example, in the Sprint network, the time between two successive optical failures ranges from 5.5 sec to 7.5 days with a mean of 12 hours [6]. Most of them are repaired automatically within several minutes, but those requiring human intervention (e.g., after a fiber cut) may last hours or days. It is quite probable that during that period another physical failures occur.

We have already discussed the multiple failures, or more specifically double-link physical failures, in [1]. However, the preliminary results described in [1] were not supported by any theoretical analysis, which limited our approach to an efficient heuristic only. Here we close this gap by studying a new, more general definition of survivability: If the logical topology remains connected after a failure of any $k$ physical links, then the underlying mapping is called "$k$–survivable." Consequently, a version of the SMART algorithm that finds a

$k$–survivable mapping will be henceforth called $k$–SMART.

It is worth noting that double-link physical failures were also considered in [12], [13], [14]. But these approaches use WDM layer protection and restoration mechanisms, whereas we focus on a failure recovery at the IP layer.

The organization of this paper is the following. Section II introduces the notation and formalizes the problem. Section III gives three fundamental theorems. Section IV introduces the $k$–SMART algorithm and discusses its properties. Section V describes a possible implementation and applications of $k$–SMART. Finally, Section VI concludes the paper.

## II. NOTATION AND PROBLEM FORMULATION

For self-containedness, we give in this section the notation introduced in [2]. When necessary, we extend it to multiple failures.

### A. Generalities

We use the formal notation of graph theory, mainly based on [15]. However, we also introduce the stack of our definitions well suited to the problems we tackle. The following general notation is used:

- $\phi$ corresponds to the *physical* topology,
- $L$ corresponds to the *logical* topology,
- $C$ corresponds to the *contracted* topology (introduced later in Section II-C),
- $a, b, c, d, e \ldots$ are used to denote edges/links,[1]
- $u, v, w \ldots$ are used to denote vertices/nodes,[2]
- $p$ is used to denote a path, i.e., a sequence of edges, where two consecutive edges have a common end-node. We say that a node $u$ is in a path $p$, $u \in p$, if $u$ is an end-node of at least one edge in $p$. A path $p$ from vertex $v$ to vertex $u$ will be denoted by $p_{v,u}$.

Physical and logical topologies are represented by undirected simple graphs: $G^\phi = (V, E^\phi)$ and $G^L = (V, E^L)$, respectively. $V$ is the set of vertices, $E^\phi$ and $E^L$ are the sets of undirected edges. In reality, not every physical node (i.e., optical switch) has an IP routing capability, which would imply $V^\phi \supseteq V^L$. All the the results in this paper hold for $V^\phi \supseteq V^L$, but for the sake of simplicity we have chosen to keep $V^\phi$ and $V^L$ identical ($V^\phi \equiv V^L \equiv V$).

### B. Lightpath and mapping

*Definition 1 (Lightpath):* A logical link $e^L$ is mapped on a physical topology as a physical path $p^\phi$ in such a way that $p^\phi$ connects the same two vertices in $G^\phi$ as $e^L$ connects in $G^L$.

In optical networking terminology, such a physical path $p^\phi$ is called a *lightpath*. The failure of any physical link in $p^\phi$ breaks the lightpath and consequently brings down the logical link $e^L$. Note that, since we release the capacity constraints, we do not have to consider the wavelengths assigned to lightpaths and wavelength converters placement.

[1] The terms *edge* and *link* will be used interchangeably
[2] The terms *vertex* and *node* will be used interchangeably

*Definition 2 (Mapping):* Let $P^\phi$ be a set of all possible physical paths in the physical topology and $A \subset E^L$ be a set of logical links. A *mapping* $M_A$ is a function $M_A : A \to P^\phi$ associating each logical link from the set $A$ with a corresponding lightpath in the physical topology.

For some particular logical edge $e^L \in A$, $M_A$ returns a physical path $p^\phi = M_A(e^L)$, $p^\phi \in P^\phi$. For arguments beyond $A$, $M_A$ is not defined. We also allow putting a set of logical links $A_{sub} \subset A$ as an argument, which results in a set of lightpaths $M_A(A_{sub}) \subset P^\phi$. Similarly, taking as an argument a logical path $p^L$ whose edges are in $A$, we obtain a set of lightpaths $M_A(p^L) \subset P^\phi$ associated with the edges of $p^L$.

*Example 1:* Fig. 1 illustrates the definitions given above. In Fig. 1a the mapping $M_A$ is defined for the subset $A$ of logical links (marked in bold in the logical topology). For example, we have $M_A(f^L) = \langle d^\phi, b^\phi, g^\phi \rangle$, which means that the lightpath assigned for the logical edge $f^L$ consists of three physical links. Fig. 1b presents a mapping defined for the subset $B$, whereas the mapping $M_{E^L}$ in Fig. 1c is defined for all links of the logical topology $E^L = A \cup B$.

We will often deal with mappings of different subsets of logical edges. Let $A_1, A_2 \subset E^L$. For consistency, we always require that:

$$\text{for every } e^L \in A_1 \cap A_2 : \ M_{A_1}(e^L) = M_{A_2}(e^L). \quad (1)$$

The mappings $M_{A_1}$ and $M_{A_2}$ can be merged, resulting in a mapping $M_{A_3}$ defined as follows

$$A_3 = A_1 \cup A_2 \quad (2)$$
$$M_{A_3}(A_3) = M_{A_1}(A_1) \cup M_{A_2}(A_2). \quad (3)$$

For convenience of notation, we will write (2) and (3) as $M_{A_3} = M_{A_1} \cup M_{A_2}$.

### C. Contraction and Origin

In the paper we will often use the graph operator of *contraction*, which is illustrated in Fig. 2 and is defined as follows:

*Definition 3 (Contraction [15]):* *Contracting* an edge $e \in E$ of a graph $G = (V, E)$ consists in deleting that edge and merging its end-nodes into a single node. The result is called the *contraction of a graph $G$ on an edge $e$* (or simply a *contracted graph*), and is denoted by $G^C = G \downarrow e$.

By extension, we also allow contracting a set of edges $A \subset E$, resulting in a contracted graph $G^C = G \downarrow A$, obtained by successively contracting the graph $G$ on every edge of $A$. It is easy to show that the order in which the edges of $A$ are taken to contraction, does not affect the final result.

Let $G = (V, E)$, $A \subset E$ and $G^C = (V^C, E^C) = G \downarrow A$. Note that by construction $E^C = E \setminus A$. Therefore each edge of $G^C$ can be found in $G$, as depicted in Fig. 2. This is not always true for vertices. A vertex of $V^C$ may either 'originate' from a single vertex in $G$ (like $w^C$ in Fig. 2), or from a connected subgraph of $G$ (like $v^C$ and $u^C$). We call this relation an $Origin(\cdot)$.

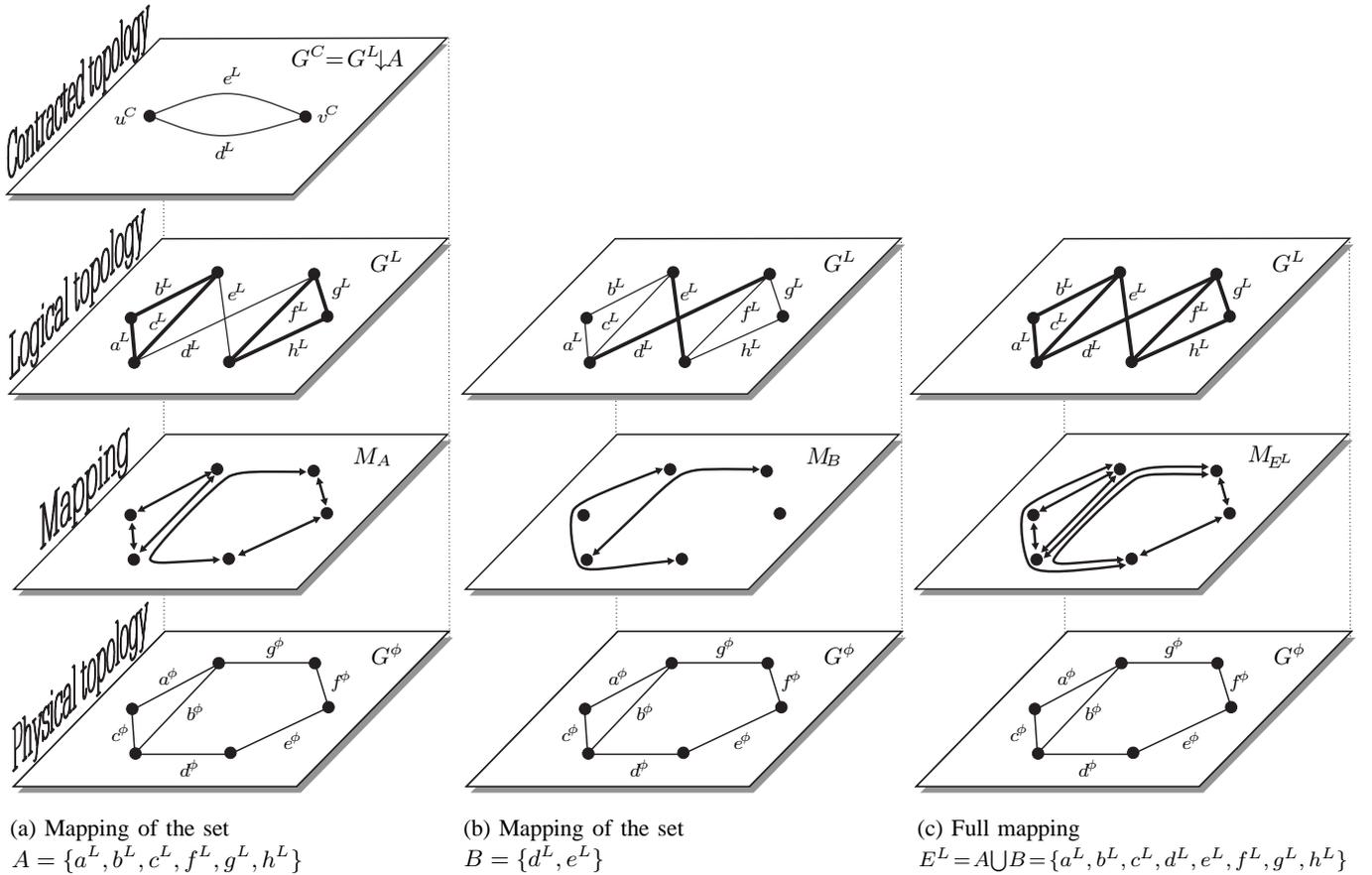

(a) Mapping of the set
$A = \{a^L, b^L, c^L, f^L, g^L, h^L\}$

(b) Mapping of the set
$B = \{d^L, e^L\}$

(c) Full mapping
$E^L = A \bigcup B = \{a^L, b^L, c^L, d^L, e^L, f^L, g^L, h^L\}$

Fig. 1. Three mapping examples. We have four layers, from bottom to top: the physical topology $G^\phi$, the mapping $M$, the logical topology $G^L$ and the contracted logical topology $G^C$ (only in (a)). In (a) the pairs $\left[G^L_{\{a^L,b^L,c^L\}}, M_A\right]$ and $\left[G^L_{\{f^L,g^L,h^L\}}, M_A\right]$ are 1–survivable, and therefore the pair $\left[G^L, M_A\right]$ is piecewise 1–survivable. In (b) the mapping $M_B$ maps edge-disjointly the set $B = \{d^L, e^L\}$ of two logical links. The contracted topology $G^C$ in (a) is composed of these two links. Taking $G^C$ and $M_B$ together, we obtain the pair $\left[G^C, M_B\right]$, which is 1–survivable. In (c) the pair $\left[G^L, M_{E^L}\right]$ is 1–survivable, that is $M_{E^L}$ is a 1–survivable mapping of the entire logical topology.

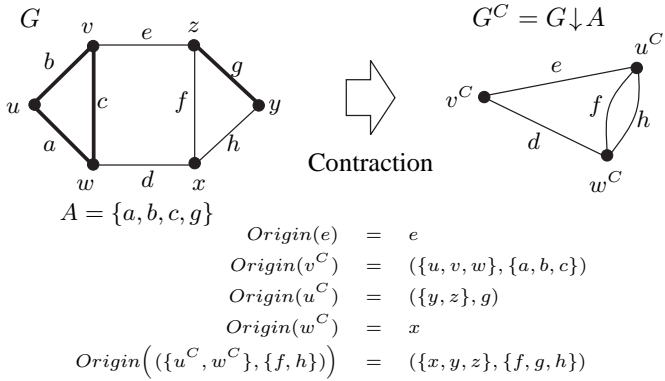

$A = \{a, b, c, g\}$

$$\begin{aligned} Origin(e) &= e \\ Origin(v^C) &= (\{u,v,w\}, \{a,b,c\}) \\ Origin(u^C) &= (\{y,z\}, g) \\ Origin(w^C) &= x \\ Origin\big((\{u^C, w^C\}, \{f, h\})\big) &= (\{x,y,z\}, \{f,g,h\}) \end{aligned}$$

Fig. 2. Contraction of a graph $G$ on a set of edges $A = \{a, b, c, g\}$. The origins of some elements of $G^C = G \downarrow A$ are also shown (bottom).

*Definition 4 (Origin):* Let $G^C = G \downarrow A$. Now take a subgraph $G^C_{sub} \subseteq G^C$. We say that $G_{sub} = Origin(G^C_{sub})$, if $G_{sub}$ is the maximal subgraph of $G$ that was transformed into $G^C_{sub}$ by the contraction of $A$ in $G$.

According to this definition, the result of the $Origin(\cdot)$ function is the *maximal* subgraph transformed in its argument.

For example, one could say that in Fig. 2, the vertex $z \in G$ was transformed into the vertex $u^C \in G^C$, however $z \neq Origin(u^C)$ because it is not the only element that was transformed into $u^C$ by contraction. The maximal subgraph in this case is $(\{y, z\}, g) = Origin(u^C)$.

### D. k–survivability and piecewise k–survivability

Let $M_{E^L}$ be a mapping of the logical topology $G^L$ on the physical topology $G^\phi$. Assume that a physical link $e^\phi$ fails. Each logical link in $G^L$ using $e^\phi$ in its mapping (lightpath) will than be cut. This may cause a disconnection of $G^L$. If, after any single physical link failure, the graph $G^L$ remains connected, then the pair $\left[G^L, M_{E^L}\right]$ is declared *1-survivable*. We extend this property to multiple failures and to a family of graphs constructed from the logical topology in the following definition:

*Definition 5 (k–survivability):* Let $G^L = (V, E^L)$, $A \subset E^L$ and $G^C = (V^C, E^C) = G^L \downarrow A$. Take any connected subgraph $G^C_{sub} = (V^C_{sub}, B)$ of the contracted topology $G^C$, and let $M_B$ be a mapping of the set $B$ of logical links. The pair $\left[G^C_{sub}, M_B\right]$ is *k–survivable* if any simultaneous failure of $k$ physical links does not disconnect the graph $G^C_{sub}$.

(Clearly, when we speak of a $k$–survivable pair, we implicitly assume the existence of a particular physical and a logical topology.)

A direct consequence of Definition 5 is that if $[G_{sub}^C, M_B]$ is $k$–survivable, then $[G_{sub}^C, M_{B'}]$ is also $k$–survivable, for any $B \subset B' \subseteq E^L$.

In Definition 5, $G_{sub}^C$ represents a large family of graphs obtained from the logical topology. If $A = \emptyset$, then $G^C = G^L$ and $G_{sub}^C$ is any connected subgraph of $G^L$ (including $G^L$ itself). If $A \neq \emptyset$, then $G_{sub}^C$ is any connected subgraph of $G^L \downarrow A$. The different instances of $G_{sub}^C$ and survivable pairs are given in Fig. 1 and described in the following three examples:

*Example 2:* One can check that in Fig. 1c the pair $[G^L, M_{E^L}]$ is 1–survivable.

*Example 3:* In Fig. 1a, let $G_{\{a^L,b^L,c^L\}}^L$ be the subgraph of $G^L$ defined by the edges $a^L, b^L, c^L$ and their end-vertices. The pair $[G_{\{a^L,b^L,c^L\}}^L, M_A]$ is 1–survivable, because a failure of any single physical link does not disconnect $G_{\{a^L,b^L,c^L\}}^L$. Similarly, the pair $[G_{\{f^L,g^L,h^L\}}^L, M_A]$ is also 1–survivable.

*Example 4:* In Fig. 1a, the contracted topology $G^C$ is the result of the contraction of the logical topology on the set $A$, i.e., $G^C = G^L \downarrow A$. Take $G_{sub}^C = G^C$. It consists of two logical links, $d^L$ and $e^L$. A possible mapping of the set $B = \{d^L, e^L\}$ is the mapping $M_B$ shown in Fig 1b. Consider the pair $[G^C, M_B]$; it is 1–survivable, because a single physical link failure cannot bring down both $d^L$ and $e^L$ at the same time, hence $G^C$ remains connected.

*Definition 6 (Piecewise $k$–survivability):* Let $M_A$ be a mapping of a set $A \subset E^L$ on the physical topology. The pair $[G^L, M_A]$ is *piecewise $k$–survivable* if, for every vertex $v^C$ of the contracted logical topology $G^L \downarrow A$, the pair $[Origin(v^C), M_A]$ is $k$–survivable.

Unlike $k$–survivability, piecewise $k$–survivability is defined only for the entire logical topology $G^L$. We will say that a mapping $M_A$ is (piecewise) $k$–survivable, if the pair $[G^L, M_A]$ is (piecewise) $k$–survivable (i.e., we take $G^L$ as the default topology).

*Example 5:* In Fig. 1a, the pair $[G^L, M_A]$ is piecewise 1–survivable. To prove it, we have to show that for vertices $u^C$ and $v^C$ of $G^L \downarrow A$, the pairs $[Origin(u^C), M_A]$ and $[Origin(v^C), M_A]$ are 1–survivable. Here we have $Origin(u^C) = G_{\{a^L,b^L,c^L\}}^L$ and $Origin(v^C) = G_{\{f^L,g^L,h^L\}}^L$. We have shown in Example 3, that each of these two graphs forms a 1–survivable pair with $M_A$.

Definition 5 can be also restated as follows:

*Definition 7 ($k$–survivability new):* Let $G^L = (V, E^L)$, $A \subset E^L$ and $G^C = (V^C, E^C) = G^L \downarrow A$. Take any connected subgraph $G_{sub}^C = (V_{sub}^C, B)$, $B \subseteq E^C$, of the contracted topology $G^C$, and let $M_B$ be a mapping of the set $B$ of logical links. The pair $[G_{sub}^C, M_B]$ is $k$–survivable if for any set $E_k^\phi \subset E^\phi$ of $k$ physical links and for any two vertices $u, v \in V_{sub}^C$, there exists a path $p_{u,v}^C$ in $G_{sub}^C$ between vertices $u$ and $v$, such that $M_B(p_{u,v}^C) \cap E_k^\phi = \emptyset$.

(Note that every path in the contracted topology, e.g., $p_{u,v}^C$, actually consists of logical links.)

In other words, a mapping is $k$–survivable if after a deletion of any set $E_k^\phi$ of $k$ physical links we can still find a path between every pair of vertices in $G_{sub}^C$. Clearly, this is equivalent to keeping $G_{sub}^C$ connected (as in Definition 5); the latter formulation is easier to be applied in the proofs in the reminder of this paper.

## III. Fundamental properties of $k$–survivable and piecewise $k$–survivable mappings

In this section we prove three useful properties of $k$–survivable and piecewise $k$–survivable mappings. We will often use them in the following sections.

### A. The expansion of $k$–survivability

Given a piecewise $k$–survivable mapping, the logical topology can be viewed as a set of $k$–survivable 'pieces'. This is a general property of a piecewise $k$–survivable mapping. (For instance in Example 5, given the piecewise 1–survivable mapping $M_A$, there are two 1–survivable 'pieces' of $G^L$: $G_{\{a^L,b^L,c^L\}}^L \subset G^L$ and $G_{\{f^L,g^L,h^L\}}^L \subset G^L$.) The following theorem enables us to merge some of these pieces, resulting in a single large $k$–survivable piece.

*Theorem 1 (Expansion of $k$–survivability):* Let $M_A$ be a mapping of a set of logical edges $A \subset E^L$ on the physical topology $G^\phi$, such that the pair $[G^L, M_A]$ is piecewise $k$–survivable. Let $G^C = G^L \downarrow A$. Take any subgraph of $G^C$, call it $G_{sub}^C = (V_{sub}^C, B)$. Let $M_B$ be a mapping of the set $B$ of edges of $G_{sub}^C$ on $G^\phi$. If the pair $[G_{sub}^C, M_B]$ is $k$–survivable then the pair $[Origin(G_{sub}^C), M_A \cup M_B]$ is also $k$–survivable.

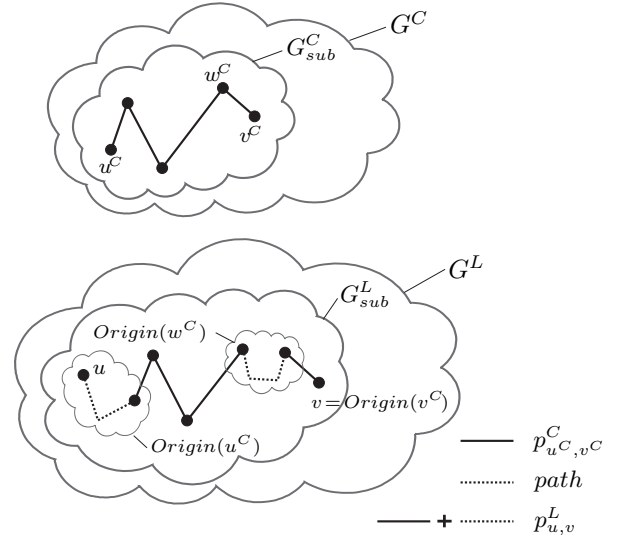

Fig. 3. Illustration of proof of Theorem 1. A first portion of the path $p_{u,v}^L$ is the path $p_{u^C,v^C}^C$ found in $G_{sub}^C$. Next it is completed, where necessary, with the patches found in origins of the nodes of $p_{u^C,v^C}^C$.

*Proof:* [Please refer to Fig. 3.]
First note that since $G^C = G^L \downarrow A$, no logical edge from the set $A$ can be found in $G^C$, which implies that $A \cap B = \emptyset$. Therefore the operation $M_A \cup M_B$ is always well defined, as in (2) and (3).

Let $M_{A \cup B} = M_A \cup M_B$ and $G_{sub}^L = Origin(G_{sub}^C)$. We have to prove that the pair $[G_{sub}^L, M_{A \cup B}]$ is $k$–survivable. Take any set $E_k^\phi \subset E^\phi$ of $k$ physical links and any two vertices $u, v \in G_{sub}^L$. According to Definition 7 we have to show that there exists a path $p_{u,v}^L$ in $G_{sub}^L$ such that $M_{A \cup B}(p_{u,v}^L) \cap E_k^\phi = \emptyset$. The path $p_{u,v}^L$ is constructed in two steps, (i) and (ii).

(i) A first portion of $p_{u,v}^L$ is found in the contracted graph $G^C$ (recall that $G^C$ consists of logical edges), as follows. Call $u^C, v^C \in V_{sub}^C$ the vertices in $G_{sub}^C = (V_{sub}^C, B)$ whose origins contain $u$ and $v$, respectively, i.e., such that $u \in Origin(u^C)$ and $v \in Origin(v^C)$. Find a path $p_{u^C, v^C}^C$ in $G_{sub}^C$, such that $M_B(p_{u^C, v^C}^C) \cap E_k^\phi = \emptyset$. This is always possible since the pair $[G_{sub}^C, M_B]$ is $k$–survivable. We take $p_{u^C, v^C}^C$ as the first portion of $p_{u,v}^L$.

(ii) We now turn our attention to the origins of vertices in the path $p_{u^C, v^C}^C$. Take any two consecutive edges $a^L$ and $b^L$ of $p_{u^C, v^C}^C$, and let $w^C$ be their common end–node in $G_{sub}^C$. If $Origin(w^C)$ is not a single node in $G_{sub}^L$, then $a^L$ and $b^L$ might not have a common end–node in $G_{sub}^L$. However, by piecewise $k$–survivability of $[G^L, M_A]$, the pair $[Origin(w^C), M_A]$ is $k$–survivable. Therefore, if we denote respectively by $v_a, v_b \in Origin(w^C)$ the end–nodes of $a^L$ and $b^L$, that belong to $Origin(w^C)$, we can find a logical path $p_{v_a, v_b}^L$ in $Origin(w^C)$ connecting $v_a$ and $v_b$, such that $M_A(p_{v_a, v_b}^L) \cap E_k^\phi = \emptyset$. We call this path a 'patch' of $w^C$ and denote it by $patch(w^C)$. If for a given $w^C$, the edges $a^L$ and $b^L$ have a common end–node $v^L$ in $G_{sub}^L$ then $patch(w^C) = v^L$.
For every vertex $w^C \in p_{u^C, v^C}^C$, find $patch(w^C)$. If $w^C = u^C$ then $patch(u^C)$ will connect the logical vertex $u$ with an end–node of the first logical edge in $p_{u^C, v^C}^C$, instead of connecting two end–nodes. The same holds for $w^C = v^C$.

To summarize, in step (i) we have found the path $p_{u^C, v^C}^C$ in the contracted subgraph $G_{sub}^C$. Next, in step (ii), we have constructed a set of *patches* for each vertex of this path. Now we combine steps (i) and (ii) to obtain the full path $p_{u,v}^L$:

$$p_{u,v}^L = p_{u^C, v^C}^C \cup \left\{ \bigcup_{w^C \in p_{u^C, v^C}^C} patch(w^C) \right\}. \quad (5)$$

The logical path $p_{u,v}^L$ connects the vertices $u$ and $v$ and has been constructed in such a way, that

$$M_B(p_{u^C, v^C}^C) \cap E_k^\phi = \emptyset \quad (6)$$
$$M_A(patch(w^C)) \cap E_k^\phi = \emptyset \quad \text{for every } w^C \in p_{u^C, v^C}^C. \quad (7)$$

Since $M_A \cup M_B = M_{A \cup B}$ and $A \cap B = \emptyset$, we can rewrite (6) and (7) as

$$M_{A \cup B}(p_{u^C, v^C}^C) \cap E_k^\phi = \emptyset \quad (8)$$
$$M_{A \cup B}(patch(w^C)) \cap E_k^\phi = \emptyset \quad \text{for every } w^C \in p_{u^C, v^C}^C. \quad (9)$$

Combining (5), (8) and (9) yields finally that $M_{A \cup B}(p_{u,v}^L) \cap E_k^\phi = \emptyset$, which proves the claim. ∎

The following example illustrates Theorem 1.

*Example 6:* In Example 5 we have shown that in Fig. 1a, the pair $[G^L, M_A]$ is piecewise 1–survivable. Take $G_{sub}^C = G^C = G^L \downarrow A$ and take $M_B$ as in Fig. 1b. From Example 4, we know that the pair $[G^C, M_B]$ is 1–survivable. Now, by Theorem 1, the pair $[Origin(G^C), M_A \cup M_B] = [G^L, M_A \cup M_B]$ is 1–survivable. So starting from the piecewise 1–survivable mapping $M_A$ and adding the mapping $M_B$, we merged the two 1–survivable pieces $G_{\{a^L, b^L, c^L\}}^L$ and $G_{\{f^L, g^L, h^L\}}^L$ into a single, large, 1–survivable piece. In this example the resulting 1–survivable piece is the entire logical topology $G^L$. The full mapping $M_A \cup M_B = M_{E^L}$ is shown in Fig. 1c.

### B. Invariance of survivability under contraction

*Theorem 2 (Invariance of $k$–survivability under contraction):* Let $G_{sub}^C = (V_{sub}^C, B)$ be a subgraph of some contracted topology $G^C$. If $M_B$ is a mapping such that the pair $[G_{sub}^C, M_B]$ is $k$–survivable, then for any set $A \subset B$ of logical links the pair $[G_{sub}^C \downarrow A, M_B]$ is also $k$–survivable.

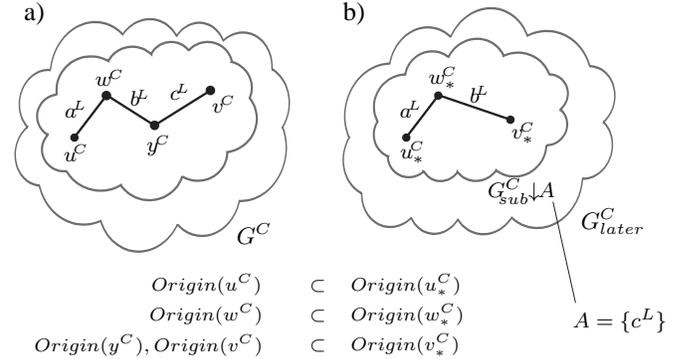

$Origin(u^C) \subset Origin(u_*^C)$
$Origin(w^C) \subset Origin(w_*^C)$
$Origin(y^C), Origin(v^C) \subset Origin(v_*^C)$

$A = \{c^L\}$

Fig. 4. Illustration of the proof of Theorem 2. (a) The original subgraph $G_{sub}^C$ and a path $p_{u^C, v^C}^C$ that avoids the set of physical links $E_k^\phi$ in its mapping. (b) The subgraph $G_{sub}^C$ contracted on the set $A = \{c^L\}$ of logical edges; the resulting subgraph is denoted by $G_{sub}^C \downarrow A$. The path $p_{u_*^C, v_*^C}^C$ originates from $p_{u^C, v^C}^C$, hence it also avoids $E_k^\phi$ in its mapping.

*Proof:* [Please refer to Fig. 4]
Take any set $E_k^\phi \subset E^\phi$ of $k$ physical links and any two vertices $u_*^C, v_*^C \in G_{sub}^C \downarrow A$. According to Definition 7 we have to show that there exists a path $p_{u_*^C, v_*^C}^C$ in $G_{sub}^C \downarrow A$ such that $M_B(p_{u_*^C, v_*^C}^C) \cap E_k^\phi = \emptyset$.
First, find in $G_{sub}^C$ two vertices $u^C, v^C \in V_{sub}^C$, such that

$$Origin(u^C) \subseteq Origin(u_*^C), \text{ and} \quad (10)$$
$$Origin(v^C) \subseteq Origin(v_*^C). \quad (11)$$

Note that since $G_{sub}^C \downarrow A$ is created by contracting some edges in $G_{sub}^C$, vertices $u^C$ and $v^C$ always exist (they are not necessarily unique). Since the pair $[G_{sub}^C, M_B]$ is $k$–survivable, there exists a path $p_{u^C, v^C}^C$ in $G_{sub}^C$ such that $M_B(p_{u^C, v^C}^C) \cap E_k^\phi = \emptyset$. Define a sequence of logical edges $p_*^C$ by contracting in $p_{u^C, v^C}^C$ all edges that exist also in $A$, i.e.,

$$p_*^C = p_{u^C, v^C}^C \downarrow (A \cap p_{u^C, v^C}^C). \quad (12)$$

Since $p_{u^C,v^C}^C$ is a path in $G_{sub}^C$, and since the contraction an edge merges its two end-nodes and thus preserves its continuity, $p_*^C$ is a path in $G_{sub}^C {\downarrow} A$. Moreover, the relations (10,11) imply that the path $p_*^C$ connects $u_*^C$ and $v_*^C$ in $G_{sub}^C {\downarrow} A$. Finally, $M_B(p_{u^C,v^C}^C) \cap E_k^\phi = \emptyset$ and (12) yields that $M_B(p_*^C) \cap E_k^\phi = \emptyset$. Therefore $p_*^C$ is the path $p_{u_*^C,v_*^C}^C$ that we are searching for. ∎

In other words, Theorem 2 says that if we can map in a $k$–survivable way some subgraph $G_{sub}^C$ of the logical or contracted logical topology, then the subgraph obtained by contracting some additional set $A$ of edges can always be mapped in a $k$–survivable way, whatever the choice of $A$.

*Example 7:* Take $G_{sub}^C = G^L$ and $M_B = M_{E^L}$ as in Fig. 1c. We know that the pair $[G^L, M_{E^L}]$ is 1–survivable. Theorem 2 implies that for any set of logical edges $A \subset E^L$ the pair $[G^L {\downarrow} A, M_{E^L}]$ is also 1–survivable. In particular, for the set $A$ as defined in Fig. 1a, $[G^L {\downarrow} A, M_{E^L}]$ is 1–survivable, which was shown in Example 4 ($M_B \subset M_{E^L}$).

Note that we do not impose any requirements (such as e.g., preserving piecewise $k$–survivability) on the contracted edges $A$. Moreover, we do not have any restrictions on what happens with the rest of the contracted topology, i.e., in $G^C \setminus G_{sub}^C$.

### C. The existence of a $k$–survivable mapping

In general, for a given pair of physical and logical topologies, it is very difficult to verify the existence of a $k$–survivable mapping. A heuristic approach, if fails, does not give any answer. The ILP approach or an exhaustive search could provide us with the answer, but due to their high computational complexity their application is limited to the topologies of several nodes. The following theorem shows how this verification problem can be substantially reduced:

*Theorem 3 (Existence of a $k$–survivable mapping):* Let $M_A$ be a mapping of a set of logical edges $A \subset E^L$, such that the pair $[G^L, M_A]$ is piecewise $k$–survivable. A $k$–survivable mapping $M_{E^L}^{surv}$ of $G^L$ on $G^\phi$ exists if and only if there exists a mapping $M_{E^L \setminus A}^{surv}$ of the set of logical links $E^L \setminus A$ on $G^\phi$, such that the pair $[G^L {\downarrow} A, M_{E^L \setminus A}^{surv}]$ is $k$–survivable.

*Proof:*
$\Leftarrow$ We know that the pair $[G^L, M_A]$ is piecewise $k$–survivable. Suppose that there exists a mapping $M_{E^L \setminus A}^{surv}$, such that the pair $[G^L {\downarrow} A, M_{E^L \setminus A}^{surv}]$ is $k$–survivable. Then, by Theorem 1, the pair $[Origin(G^L {\downarrow} A), M_A \cup M_{E^L \setminus A}^{surv}] = [G^L, M_A \cup M_{E^L \setminus A}^{surv}]$ is also $k$–survivable. So the mapping $M_{E^L}^{surv} = M_A \cup M_{E^L \setminus A}^{surv}$ is a $k$–survivable mapping of $G^L$ on $G^\phi$.
$\Rightarrow$ Assume that a $k$–survivable mapping of $G^L$ on $G^\phi$ exists, call it $M_{E^L}^{surv}$. Now, by taking $G_{sub}^C := G^L$ and $M_B := M_{E^L}^{surv}$, Theorem 2 yields that $[G^L {\downarrow} A, M_{E^L}^{surv}]$ is $k$–survivable. Consequently, the pair $[G^L {\downarrow} A, M_{E^L \setminus A}^{surv}]$ is also $k$–survivable. ∎

The following example illustrates this theorem.

*Example 8:* In Fig. 1 delete edge $b^\phi$ from the physical topology $G^\phi$. Now, for the logical topology $G^L$ and the physical topology $G^\phi \setminus \{b^\phi\}$, a 1–survivable mapping does not exist. To prove it, note that we can still easily find a mapping $M_A$ of $G^L$ on $G^\phi \setminus \{b^\phi\}$ that is piecewise 1–survivable. However, the remaining two logical links $d^L$ and $e^L$, cannot be mapped edge-disjointly on $G^\phi \setminus \{b^\phi\}$. Therefore no 1–survivable mapping $M_{\{d^L, e^L\}}$ of the contracted logical topology $G^L {\downarrow} A$ on $G^\phi \setminus \{b^\phi\}$ exists. Consequently, by Theorem 3 we know that no 1–survivable mapping of $G^L$ on $G^\phi \setminus \{b^\phi\}$ exists, which was to be proved. Note that to prove it we only considered the two-edge topology $G^L {\downarrow} A$ instead of the entire $G^L$, which greatly simplified the problem. Clearly, the larger the set $A$, the more we benefit from Theorem 3.

## IV. THE $k$–SMART ALGORITHM

In this section we present an algorithm that searches for a $k$–survivable mapping. We call this algorithm $k$–SMART, as it is a straightforward extension of the SMART algorithm [1], [2] to multiple failures. It maps the topology part by part, gradually converging to a final solution. By formal graph theoretic analysis, we prove that if $k$–SMART converges completely, a *$k$–survivable* mapping is found. Otherwise, when the algorithm terminates before its complete convergence, the returned mapping is *piecewise $k$–survivable* and no $k$–survivable solution exists.

### A. The pseudo-code of $k$–SMART

*Step 1* Start from the full logical topology $G^C = G^L$, and an empty mapping $M_A = \emptyset$, $A = \emptyset$;
*Step 2* Take some subgraph $G_{sub}^C = (V_{sub}^C, B)$ of $G^C$ and find a mapping $M_B$, such that the pair $[G_{sub}^C, M_B]$ is $k$–survivable. IF no such pair exists, THEN RETURN $M_A$ AND $G^C = G^L {\downarrow} A$, END.
*Step 3* Update the mapping by merging $M_A$ and $M_B$, i.e., $M_A := M_A \cup M_B$;
*Step 4* Contract $G^C$ on $B$, i.e., $G^C := G^C {\downarrow} B$;
*Step 5* IF $G^C$ is a single node, THEN RETURN $M_A$, END.
*Step 6* GOTO Step 2

The $k$–SMART algorithm starts from an empty mapping $M_A = \emptyset$. At each iteration it maps some set $B$ of logical links (Step 2), and extends the mapping $M_A$ by $M_B$ (Step 3). Meanwhile, the contracted topology $G^C$ gradually shrinks (Step 4).

### B. The correctness of the $k$–SMART algorithm

We will declare that:
• *$k$–SMART converges* if the contracted topology $G^C$ converges to a *single* node. We prove later in Corollary 1, that the mapping $M_A$ returned in step 5 is then a $k$–survivable solution;
• *$k$–SMART does not converge* if $k$–SMART terminates before $G^C$ converges to a single node. This happens when Step 2 of $k$–SMART is impossible to make. We prove below in Theorem 4 that the mapping $M_A$ returned in Step 2 piecewise $k$–survivable. Moreover, we show in Corollary 1 that in this case a $k$–survivable solution does not exist. The graph $G^C = G^L \downarrow A$ (also returned in Step 2) we call the *remaining contracted logical topology* since it consists of unmapped logical links $E^L \backslash A$.

*Theorem 4 ($k$–SMART's piecewise $k$–survivability):* After each iteration of the $k$–SMART algorithm, the pair $[G^L, M_A]$ is piecewise $k$–survivable.

*Proof:* [By induction]
INITIALIZATION:
Initially $G^C = G^L$. Therefore the origin of any vertex $v^C \in V^C$ is a single node in $G^L$, and it cannot be disconnected. Hence for every $v^C \in V^C$, the pair $[Origin(v^C), M_A]$ is $k$–survivable and consequently the pair $[G^L, M_A]$ is piecewise $k$–survivable.
INDUCTION:
Assume that after some iteration the pair $[G^L, M_A]$ is piecewise $k$–survivable. We have to prove that after the next iteration of the algorithm, the updated mapping $\widehat{M_A}$ will still form a piecewise $k$–survivable pair $[G^L, \widehat{M_A}]$.
One iteration of the $k$–SMART algorithm consists of Steps 2, 3 and 4, which we recall here:
2. Find $G^C_{sub} = (V^C_{sub}, B)$ and $M_B$, such that the pair $[G^C_{sub}, M_B]$ is $k$–survivable.
3. $\widehat{M_A} := M_A \cup M_B$
4. $\widehat{G^C} := G^C \downarrow B$
(For clarity we indicated the updated $M_A$ and $G^C$ by a hat: "$\widehat{\phantom{x}}$")
The updated contracted topology $\widehat{G^C} = (\widehat{V^C}, \widehat{E^C})$ was created from $G^C$ by replacing $G^C_{sub} = (V^C_{sub}, B)$ with a single node, which we call $\widehat{v}^C_{sub}$; the remaining nodes stayed unchanged. So $\widehat{V^C} = \{\widehat{v}^C_{sub}\} \cup V^C \backslash V^C_{sub}$. Take any $\widehat{v}^C \in \widehat{V^C}$; we have two possibilities:
(i) $\widehat{v}^C = \widehat{v}^C_{sub}$: Since $G^C_{sub} = (V^C_{sub}, B)$ was contracted into $\widehat{v}^C_{sub}$, their origins coincide: $Origin(G^C_{sub}) = Origin(\widehat{v}^C_{sub})$. Since $\widehat{M_A} = M_A \cup M_B$, the pair $[Origin(\widehat{v}^C_{sub}), \widehat{M_A}] = [Origin(G^C_{sub}), M_A \cup M_B]$ is $k$–survivable by Theorem 1.
(ii) $\widehat{v}^C \neq \widehat{v}^C_{sub}$: In this case $\widehat{v}^C \in V^C \backslash V^C_{sub}$, so $\widehat{v}^C = v^C$. By piecewise $k$–survivability of the pair $[G^L, M_A]$, the pair $[Origin(v^C = \widehat{v}^C), M_A]$ is $k$–survivable. Since $\widehat{M_A} = M_A \cup M_B$, the pair $[Origin(\widehat{v}^C), \widehat{M_A}]$ is $k$–survivable as well.
Combining (i) and (ii), we have proven that for every $\widehat{v}^C \in \widehat{V^C}$, the pair $[Origin(\widehat{v}^C), \widehat{M_A}]$ is $k$–survivable. So, by Definition 6, the pair $[G^L, \widehat{M_A}]$ is piecewise $k$–survivable. ∎

Theorem 4 leads us to the following important property of $k$–SMART:

*Corollary 1 ($k$–SMART's convergence):* The $k$–SMART algorithm returns a single node contracted topology $G^C$ if and only if there exists a $k$–survivable mapping of the logical graph $G^L$ on the physical graph $G^\phi$. In this case the returned mapping $M_A$ is $k$–survivable.

*Proof:*
$\Rightarrow$ We have to show that if there is only one vertex in $G^C$ then $[G^L, M_A]$ is $k$–survivable.
We have two observations: (i) By Theorem 4, the pair $[G^L, M_A]$ is piecewise $k$–survivable. This means that for every vertex $v^C \in G^C$ the pair $[Origin(v^C), M_A]$ is $k$–survivable. (ii) There is only one vertex in $G^C$ (i.e., $G^C = \{v^C\}$), and therefore $Origin(v^C) = G^L$. Combining (i) and (ii), we have that $[G^L, M_A]$ is $k$–survivable.
$\Leftarrow$ We have to show that if the contracted topology $G^C$ has more than one node then a $k$–survivable mapping of $G^L$ on $G^\phi$ does not exist.
By Theorem 4, the pair $[G^L, M_A]$ is piecewise $k$–survivable. Since the algorithm has returned before converging to a single node (i.e., in Step 2), there exists no pair $[G^C_{sub}, M_B]$ that is $k$–survivable. In particular, if we take $G^C_{sub} = G^C = G^L \downarrow A$, there exists no pair $[G^L \downarrow A, M_*]$ that is $k$–survivable. Now, by Theorem 3 there exists no $k$–survivable mapping of $G^L$ on $G^\phi$. ∎

$G^C$ may converge to a single node topology with *self-loops*; they form a set of remaining unmapped logical links $E^L \backslash A$. However, this does not affect the result, because the links of $E^L \backslash A$ may be mapped in any way (e.g. shortest path) to obtain a full $k$–survivable mapping $M_{E^L}$.

### C. The order of a sequence of subgraphs

Recall that in Step 2 of the $k$–SMART algorithm we take some subgraph $G^C_{sub} = (V^C_{sub}, B)$ of the contracted topology $G^C$. We do not specify which subgraph to take; if there are more candidates $G^C_{sub}$ that meet the condition given in Step 2 (which is usually the case), we are free to pick any of them. This raises a natural question: How does the choice of $G^C_{sub}$ affect the convergence of the $k$–SMART algorithm? In the following theorem we show that, in general, this choice does *not* affect the outcome of the $k$–SMART algorithm.

*Theorem 5 ($k$–SMART unique convergence):* There exists a unique contracted topology $G^C_{min}$ (excluding self-loops) returned by $k$–SMART.

*Proof:* [By contradiction, Please refer to Fig. 5]
Let us assume that two different runs of $k$–SMART converge to two different contracted topologies $G^C_1 = G^L \downarrow A$ and $G^C_2 = G^L \downarrow B$, and the mappings $M_A$ and $M_B$, respectively. The $k$–SMART algorithm returned in Step 2, which implies that no subgraph $G^C_{sub1}$ of $G^C_1$ can be mapped in a $k$–survivable way; similarly, no subgraph $G^C_{sub2}$ of $G^C_2$ can be mapped in a $k$–survivable way. Assume, without loss of generality, that there exists an edge $e^L_*$ such that $e^L_* \in G^C_2$ and $e^L_* \notin G^C_1$. (If such an edge does not exist, an edge satisfying a converse condition must exist, because $G^C_1 \neq G^C_2$.) Since $e^L_* \notin G^C_1$, there exists $v^C_* \in G^C_1$ such that $e^L_* \in Origin(v^C_*)$. By Theorem 4, the pair $[G^C_1, M_A]$ is piecewise $k$–survivable, which implies that $[Origin(v^C_*), M_A]$ is $k$–survivable. Now, by Theorem 2, the pair $[Origin(v^C_*) \downarrow B, M_A]$ is also $k$–survivable. By construction the subgraph $Origin(v^C_*) \downarrow B$ contains at least the edge $e^L_*$. Therefore, there exists a non-empty subgraph $G^C_{sub} = Origin(v^C_*) \downarrow B$ of $G^C_2$ that can be mapped in a $k$–survivable way (using the mapping $M_A$), which is impossible because no subgraph $G_{sub2}^C$ of $G_2^C$ can be mapped in a $k$–survivable way. ∎

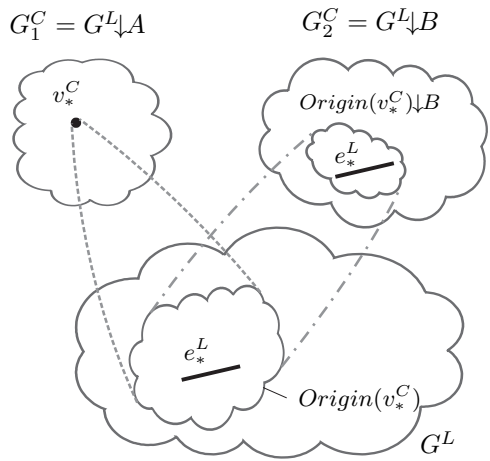

Fig. 5. Illustration of proof of Theorem 5. We start with an edge $e_*^L$ that is in $G_2^C$, but not in $G_1^C$. Next, we choose a vertex $v_*^C \in G_1^C$ such that $e_*^L \in Origin(v_*^C)$. In the topology $G_2^C$, $Origin(v_*^C)$ is contracted to $Origin(v_*^C) \downarrow B$ that contains at least $e_*^L$. This nonempty subgraph $Origin(v_*^C) \downarrow B$ can be mapped in a $k$–survivable way using the mapping $M_A$, which leads to contradiction.

A direct consequence of Theorem 5 is that the order in which we take $G_{sub}^C$ in the $k$–SMART algorithm does not affect the final result.

## V. IMPLEMENTATION AND APPLICATIONS

In practice, it is not feasible to implement the exact code given in IV-A, because Step 2 alone is an NP-complete problem. A possible practical solution is to restrict the types of subgraphs $G_{sub}^C$ taken in Step 2 of the $k$–SMART algorithm. Clearly, in order to map a graph $G_{sub}^C$ in a $k$–survivable way, $G_{sub}^C$ has to be a $(k{+}1)$–edge–connected. For instance, to achieve a 2-survivability we can consider in Step 2 the 3–edge–connected structures shown in Fig. 6. We have implemented this in [1] with very good results.

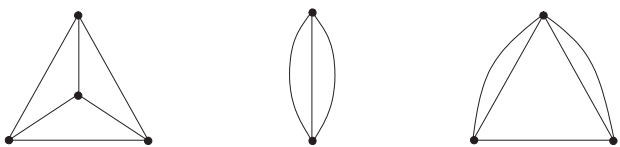

Fig. 6. Possible subgraphs $G_{sub}^C$ that can be considered in Step 2 in the implementation of the $k$–SMART algorithm, for $k = 2$.

Since we have, in this paper, extended all the theorems from [2] to multiple failure scenarios, all applications of SMART described in [2] naturally carry over to $k$–SMART. In particular, we can apply the $k$–SMART algorithm as:

- the formal verification of the existence of a $k$–survivable mapping,
- a tool tracing and repairing the vulnerable areas of the network,
- a fast heuristic.

## VI. CONCLUSIONS

In this paper we have extended all the theoretical results in [2] to the presence of multiple link failures. In the future we plan to apply these results to design a mapping robust to multiple failures in various scenarios in IP-over-WDM networks.

The work presented in this paper was financially supported by grant DICS 1830 of the Hasler Foundation, Bern, Switzerland.